\documentclass[aps,prl,twocolumn,showpacs,preprintnumbers,superscriptaddress]{revtex4-1}
\usepackage{graphicx,bm,dcolumn,amsmath,amssymb,amsfonts}
\newcommand{\vm}{\bm{m}}

\newcommand{\vn}{\bm{n}}
\newcommand{\vx}{\bm{x}}
\newcommand{\vy}{\bm{y}}
\newcommand{\vsigma}{\bm{\sigma}}

\begin{document}

\title{Geometric Quantum Gates, Composite Pulses, 
and Trotter-Suzuki Formulas} 
 
\author{Yasushi Kondo}
\affiliation{
Department of Physics, Kinki University, 
Higashi-Osaka, 577-8502, Japan}
\affiliation{
Research Center for Quantum Computing, Interdisciplinary Graduate School 
of Science and Engineering, Kinki University, \\
Higashi-Osaka, 577-8502, Japan}
\author{Masamitsu Bando}
\affiliation{
Research Center for Quantum Computing, Interdisciplinary Graduate School 
of Science and Engineering, Kinki University, \\
Higashi-Osaka, 577-8502, Japan}
\date{\today}

\begin{abstract}
We show that all geometric quantum gates (GQG's in short), 
which are quantum gates only with geometric 
phases, are robust against control field strength errors.  
As examples of this observation, we show 
(1) how robust composite \mbox{rf-pulses} in NMR are geometrically
constructed and
(2) a composite \mbox{rf-pulse} based on Trotter-Suzuki Formulas 
is a GQG. 
\end{abstract}

\maketitle

Geometric phases have been attracting a lot of attention from 
the view point of the foundation of quantum mechanics and mathematical
physics~\cite{Nakahara2003,ChruscinskiJamiolkowski2004,Vojta2003,
BengtssonZyczkowski2006}. 
Recently,  a geometric quantum gate (GQG in short), 
which is a quantum gate only with geometric phases,
is spotlighted in quantum information 
processing~\cite{others,Tianetal2004}, because they
are expected to be robust against noise. 
Although its robustness has not yet been generally
confirmed~\cite{BlaisTremblay2003,ZhuZanardi2005,NazirSpillerMunro2002,
Carollo;Vedral:2003,Chiara;Palma:2003,Dajka;Luczka:2008}, 
some GQG's are robust against certain types of 
fluctuations~\cite{Ohta2009}.   

On the other hand, composite \mbox{rf-pulses} 
are extensively employed in NMR~\cite{Levitt1986,Claridge1999}, 
which are robust against systematic errors of the system. 
Note that \mbox{rf-pulses} are means for controlling spin states 
and have direct correspondence to quantum gates. 
Most of composite \mbox{rf-pulses} in NMR are designed with the knowledge 
of initial states, and thus it is often not replaceable with simple 
pulses. However, there are {\it fully compensating}
composite \mbox{rf-pulses} that are replaceable with simple pulses
without further modifications of other pulses, and thus 
are compatible in use for quantum computation, as demonstrated in
ion traps~\cite{ion} and Josephson junctions~\cite{JJ}
as well as in NMR~\cite{Jones2000}. 

In this letter, 
we discuss the relation between fully compensating composite 
quantum gates which is robust against control field strength errors
and non-adiabatic GQG's 
with Aharanov-Anandan (AA) phases~\cite{AharonovAnandan1987}.
Let us define an ideal single-qubit operation 
\begin{eqnarray}
R(\vm, \theta) &=& \exp(- i \theta \frac{\vm \cdot \vsigma}{2}),
\label{eq:single_rotation}
\end{eqnarray}
where we take the natural unit system 
in which $\hbar =1$. $\vm$ is a unit vector ($\in \mathbb{R}^3$), 
while $\vsigma$ is a standard Pauli matrix vector such that
$\vsigma=(\sigma_{x},\,\sigma_{y},\,\sigma_{z})$.
$\theta$ represents a control field strength. Note that $\theta$ and 
$\vm$ are both constant.
A real erroneous operation $\tilde{R}(\vm, \theta)$
with a systematic control field strength 
error is modeled as follows. 
\begin{eqnarray*}
\tilde{R}(\vm, \theta) 
&=& R(\vm, \theta(1+\epsilon)) 
= R(\vm, \theta)+ O(\epsilon),
\end{eqnarray*}
where $\epsilon \ll 1$ is an unknown fixed parameter that represents 
the error. 
If we find a series of 
operations $\tilde{R}(\vm_j, \theta_j)$ ($\vm_j$ and $\theta_j$
are constant), such that
\begin{eqnarray}
\prod_{j=1}^N \tilde{R}(\vm_j, \theta_j)  &=& R(\vm, \theta)
+ O(\epsilon^2),  
\label{eq:composite_pulse}
\end{eqnarray}
we call it a fully compensating composite quantum gate 
which is robust against a control field strength error.

In order to proceed our discussion, we first review an AA phase 
that appears under non-adiabatic cyclic time evolution of a
quantum system~\cite{AharonovAnandan1987} and a GQG
with it for a single qubit~\cite{Ohta2009}.
The qubit state $|\vn(t)\rangle(\in\mathbb{C}^{2})$ at $t$ ($\in [0,T]$)
corresponds to the Bloch vector 
$\vn(t)=\langle \vn(t)| \vsigma | \vn(t)\rangle (\in\mathbb{R}^{3})$.
Suppose that the Hamiltonian $H(t)$  generates a cyclic time evolution 
such that  \(|\vn(T)\rangle = e^{i\gamma}|\vn(0)\rangle\)
($\gamma\in\mathbb{R}$).
The AA phase $\gamma_{{\rm g}}$ is defined as~\cite{AharonovAnandan1987}
\begin{equation}
\gamma_{{\rm g}} = \gamma - \gamma_{{\rm d}},
\label{eq:def_AAphase}
\end{equation}
where 
\begin{equation}
\gamma_{{\rm d}} 
= -\int^{T}_{0}\langle \vn(t)|H(t)|\vn(t)\rangle\, dt
\label{eq:dyn_phase}
\end{equation}
is a dynamic phase. 
Next, suppose $|\vn_{+}(0)\rangle$ and $|\vn_{-}(0)\rangle$ 
are two states satisfying 
\mbox{(a) $\langle \vn_{+}(0) | \vn_{-}(0) \rangle=0$} 
(or, \mbox{$\vn_{+}(0) = - \vn_{-}(0)$}) and
\mbox{(b) 
$|\vn_{\pm}(T) \rangle = e^{i \gamma_{\pm}} |\vn_{\pm}(0)\rangle$}, 
where $\gamma_{\pm} \in \mathbb{R} $.  
An arbitrary quantum state $|\vn (0)\rangle$ is expressed as
\mbox{$|\vn(0) \rangle = a_{+}|\vn_{+}(0)\rangle 
                       + a_{-}|\vn_{-}(0)\rangle$}, where 
\mbox{$a_{\pm}=\langle \vn_{\pm}(0)|\vn(0)\rangle$}. 
We call $\vn_{\pm}(0)$ as a 
basis Bloch vector associated with $H(t)$.
The initial state $|\vn(0)\rangle$ is transformed into the final state 
\begin{eqnarray*}
|\vn(T)\rangle  
&=&  a_{+}\,e^{i \gamma_{+}}|\vn_{+}(0)\rangle 
+  a_{-}\,e^{i \gamma_{-}}|\vn_{-}(0)\rangle.
\end{eqnarray*}
Thus, the time evolution operator $U$ at $t=T$ generated by $H(t)$
($t\in [0,T]$) is rewritten as
\begin{eqnarray}
 U &=& 
 e^{i\gamma_{+}}|\vn_{+}(0)\rangle\langle \vn_{+}(0)|
+
 e^{i\gamma_{-}}|\vn_{-}(0)\rangle\langle \vn_{-}(0)|.
\label{eq:geometricgate}
\end{eqnarray}
Eq.~(\ref{eq:geometricgate}) is regarded as a quantum gate with an AA
geometric phase, when the dynamic component of $\gamma_{\pm}$ is
vanishing. 

Let us discuss the following Hamiltonian 
\begin{equation}
 H(\vm, \theta) = \theta \frac{\vm \cdot \vsigma}{2}\frac{1}{T},
\label{eq:Hamiltonian}
\end{equation}
at $ t\in [0,T]$ \cite{memo}. Applying this Hamiltonian for $[0,T]$, 
$R(\vm, \theta)$ (Eq.~(\ref{eq:single_rotation})) is obtained. 
The dynamic phase $\gamma_{\rm d}$ generated 
by $H(\vm, \theta)$ at  $t \in [0,T]$ 
for the system starting from $|\vn \rangle $ is given as~\cite{Ohta2009}
\begin{eqnarray}
\gamma_{\rm d} 
&=& -\int_0^T \langle \vn| H(\vm, \theta)|\vn\rangle dt
\nonumber \\
&=& \langle \vn| H(\vm, \theta) \, T |\vn\rangle
= - \frac{\theta}{2} \,\,\vm \cdot \vn.
\label{eq:dyn_p}
\end{eqnarray}
The state $|\pm \vm\rangle$  is a cyclic state for 
$H(\vm, \theta)$ and there is no other cyclic state except for 
$|\pm \vm\rangle$ for $H(\vm, \theta)$ if 
$\theta ({\rm mod} \,\,2\pi) \ne 0$. 
By using $|\pm \vm \rangle$, $R(\vm, \theta)$ is rewritten as 
\begin{eqnarray*}
R(\vm, \theta) &=& 
   e^{-i \theta/2}| \vm \rangle \langle  \vm | 
 + e^{ i \theta/2}|-\vm \rangle \langle -\vm | ,
\end{eqnarray*}
like Eq.~(\ref{eq:geometricgate}). 
We may call $R(\vm, \theta)$
as a dynamic phase gate by contrasting a GQG.

We are ready to discuss a composite quantum gate 
\mbox{$ \prod_{j=1}^N \tilde{R}_j$}, such that  
\mbox{$ \prod_{j=1}^N R_j = R(\vn_0, \theta)$}.  
$\tilde{R}(\vm_j, \theta_j)$ and  $R(\vm_j, \theta_j)$ are  
abbreviated to $\tilde{R}_j$ and $R_j$, respectively. 
We assume that each duration of $R_j$ is 
$T_j$ and thus the associated Hamiltonian for $R_j$ is 
\mbox{$\displaystyle 
H_j = \theta_j \frac{\vm_j \cdot \vsigma}{2}\frac{1}{T_j}$}. 
By ignoring $O(\epsilon^2)$, 
\begin{eqnarray*}
&&\prod_{j=1}^N \tilde{R}(\vm_j, \theta_j) \nonumber \\
&=&\prod_{j=1}^N R(\vm_j, \theta_j(1+\epsilon)) \nonumber \\
&=& R(\vn_0, \theta)
+ \sum_{j=1}^N  R_N\dots R_j \left(R(\vm_j, \theta_j \epsilon)-I\right)
\dots R_1 \nonumber \\
&=& 
R(\vn_0, \theta) + \sum_{j=1}^N  R_N\dots R_j\left( 
I - i\epsilon \theta_j\frac{\vm_j\cdot\vsigma}{2} - I
\right) \dots R_1
\nonumber \\
&=& R(\vn_0, \theta)
- i \epsilon \sum_{j=1}^N R_N \dots R_j\left( H_j T_j\right) \dots R_1, 
\end{eqnarray*}
where $I$ is the identity operator 
for a qubit. 
We calculate the expectation value of the second term 
for $|\vn_0\rangle$ that is a cyclic state for both $R(\vn_0, \theta)$
and $ \prod_{j=1}^N R_j $ by definition.
\begin{eqnarray}
&&\langle \vn_0 |  -\sum_{j=1}^N R_N \dots R_j H_j T_j
R_{j-1}\dots R_1 |\vn_0\rangle 
\nonumber \\
&=& - e^{-i \theta/2} \sum_{j=1}^N 
\langle \vn_{j-1} | H_j T_j|\vn_{j-1} \rangle  
\label{eq:exp_v}
\end{eqnarray}
where $|\vn_j \rangle = \prod_{k=1}^j R_k |\vn_0\rangle$. Note that 
$\langle \vn_0 | R_N \dots R_j = e^{-i\theta/2}\langle \vn_{j-1}|$
by definition of $\prod_{j=1}^N R_j $.

If $\prod_{j=1}^N \tilde{R}_j$ is a fully compensating composite quantum 
gate which is robust against a control field strength error, 
$\sum_{i=1}^N  R_N \dots R_j (H_j T_j) \dots R_1$ 
should vanish by definition, or by Eq.~(\ref{eq:composite_pulse}). 
And thus, Eq.~(\ref{eq:exp_v}) should vanish. On the other hand, 
$-\langle \vn_{j-1} | H_j T_j|\vn_{j-1} \rangle $
is a dynamic phase $\gamma_{{\rm d}, j}$ accumulated during 
the $j$'th operation $R_j$. 
Then, we conclude that
\begin{eqnarray*}
\sum_{j=1}^N \gamma_{{\rm d}, j} =0,
\end{eqnarray*}
for the cyclic state $|\vn_0\rangle$. 
It is also obvious that $\prod_{j=1}^N R_j$ 
generates no dynamic phase when starting from $|-\vn_0\rangle $
where  $\langle -\vn_0 |\vn_0\rangle =0$. Therefore, 
\begin{eqnarray*}
\prod_{j=1}^N R_j = 
  e^{  i\gamma_{\rm g}}| \vn_0 \rangle\langle  \vn_0|
+ e^{- i\gamma_{\rm g}}|-\vn_0 \rangle\langle -\vn_0|,
\end{eqnarray*}
where $\gamma_{\rm g}= -\theta/2$ 
is a geometric phase.
In other words, {\it a fully compensating composite quantum gate 
without a dynamic phase is always robust against 
a control field strength error.}

We, however, note that not-all robust composite quantum gates 
are regarded as GQG's with AA-phases. 
For example, CORPSE pulses~\cite{cp_qip} are not regarded as 
GQG's since we can only define 
cyclic states that acquire dynamic phases after 
operations. These composite \mbox{rf-pulses} rely on the 
non-linearity of pulse responses.

We now discuss some concrete examples of composite quantum gates 
in terms of vanishing dynamic phases. 

{\bf No dynamic phases during a composite quantum gate}: 
We discussed 
$R(\vx, \pi/2)R(\vy, \pi)R(\vx, \pi/2)$, 
where $\vx (\vy)$ is the unit vector 
along the x (y) axis~\cite{Ohta2009}. 
This composite \mbox{rf-pulse} is very special in the
sense that no dynamic phases are generated for its cyclic states 
at any moment of operation $R_j$. 
We noticed the importance of dynamic phases in quantum 
operations~\cite{Ohta2009}  and now extend this observation so that 
the sum of dynamic phases is important.

{\bf SCROFULOUS pulse and W1 correction sequence}: 
Cummins, Llewellyn, and Jones showed the systematic method how 
to construct a composite \mbox{rf-pulses} called SCROFULOUS pulses
\begin{eqnarray}
R(\vm_1,\theta_1)R(\vm_2,\pi)R(\vm_1,\theta_1)=R(\vx,\theta),
\label{eq:SCR}
\end{eqnarray}
where $\vm_i = (\cos \phi_i, \sin \phi_i,0)$ and $\theta_1$ is an 
rotation angle of the first and third pulses~\cite{cp_qip}.  
We take $\vx$ for simplicity, but generalization from $\vx$ 
to $\vm$ in the xy-plane should be trivial. 
We discuss another way to construct them in terms of dynamic phases. 
From Eq.~(\ref{eq:SCR})~\cite{comment}, 
\begin{eqnarray*}
\langle \vx| R(\vm_1,\theta_1)R(\vm_2,\pi)R(\vm_1,\theta_1)|\vx\rangle
&=& 
e^{-i \theta/2}.
\end{eqnarray*}
Then, we obtain 
\begin{eqnarray}
\cos \theta_1 &=& \frac{\tan(\phi_1-\phi_2)}{\tan \phi_1}, 
\label{eq:SCR1} \\
\sin \frac{\theta}{2} &=& \frac{\sin(\phi_1-\phi_2)}{\sin \phi_1}
\label{eq:SCR2}.
\end{eqnarray}
From the condition that the sum of dynamic phases is
zero~\cite{comment1} and Eq.~(\ref{eq:SCR1}), we obtain 
\begin{eqnarray}
2 \theta_1 \cos (\phi_1 - \phi_2)+\pi &=& 0
\label{eq:SCR3}.
\end{eqnarray}
By solving Eqs.~(\ref{eq:SCR1}-\ref{eq:SCR3})\cite{my_cal}, 
we obtain the same results as in Ref.~\cite{cp_qip}. 

A W1 sequence~\cite{cp_qip} which corrects $R(\vx, \theta)$
is  
\begin{eqnarray*}
U_{\rm W1} &=& R(\vm_1,\pi)R(\vm_2,2\pi)R(\vm_1,\pi) = I.
\end{eqnarray*}
Here, we take  
$\phi_1 = \pm {\rm arccos}\left(-\theta/(4\pi)\right)$ and
$\phi_2 = 3\phi_1$. 
$U_{\rm W1}$ is rewritten as 
\begin{eqnarray*}
U_{\rm W1}&=&e^{i\gamma_{\rm W1}} |\vx\rangle \langle \vx| 
+ e^{-i\gamma_{{\rm W1}}}|-\vx\rangle \langle -\vx|,
\end{eqnarray*}
where 
\mbox{$\gamma_{\rm W1}=\gamma_{{\rm g, W1}}+\gamma_{{\rm d, W1}}=0$}
and $\gamma_{\rm g, W1}$$(\gamma_{\rm d, W1})$ is 
a geometric (dynamic) phase. Since
\mbox{$\gamma_{\rm d, W1} = -\gamma_{\rm g, W1} = \theta/2$},
the composite quantum gate 
\mbox{$R(\vx, \theta/2)U_{\rm W1}R(\vx, \theta/2)$}
becomes a GQG. 

{\bf Composite \mbox{rf-pulse} based on Trotter-Suzuki formula}:
Brown, Harrow, and Chuang discussed a systematic method, 
named Trotter-Suzuki method, 
to construct a composite \mbox{rf-pulses}~\cite{BHC2004}. 
We re-examine this method in terms of a dynamic phase.
Let us consider a composite quantum gate that is equivalent to 
$R(\vx, \theta)$.   Erroneous rotation 
$\tilde{R}(\vx, \theta)=R(\vx, \theta) R(\vx, \theta \epsilon)$ 
is able to be compensated when $R(\vx, \theta \epsilon)$ is
approximately 
canceled by another erroneous rotations 
\mbox{$\tilde{R}(\vm', \theta')$'s}. 
According to Trotter-Suzuki Formulas~\cite{TS_formula}, 
we can select $A_i$'s so that 
\begin{eqnarray*}
R({\bm x}, -\theta \epsilon) 
&=& \exp\left(-i (-\theta \frac{\sigma_x}{2}) \epsilon \right) \nonumber \\
&=& \prod_{i=1}^N \exp(-i A_i \epsilon) +O(\epsilon^2) 
\end{eqnarray*}
where $-\theta (\sigma_x/2)=\sum_{i=1}^N A_i$\,
\mbox{($N \ge 2$)}. 
Now, we take \mbox{$N=2$} (the smallest number) for simplicity 
and call $A_i$ as $A_\pm$. 
We require that  $|\vx \rangle$ is the cyclic state for 
$\exp(-i A_\pm)$ as well as $R(\vx, \theta)$. $A_\pm$ that 
satisfies these conditions is, for example, 
\begin{eqnarray*}
A_\pm &=& -2\pi \frac{\vm_\pm \cdot \vsigma}{2},
\end{eqnarray*}
where $\vm_\pm = \cos \phi \, \vx \pm \sin \phi \, \vy$ 
and $\phi = \cos^{-1}(\theta/4\pi)$.  
Since 
\begin{eqnarray*}
&&R(\vm_+, -2\pi(1+\epsilon)) R(\vm_-, -2\pi(1+\epsilon))
\nonumber \\
&=& \exp(-i A_+ \epsilon)\exp(-i A_- \epsilon),  
\end{eqnarray*}
a robust composite quantum gate
\begin{eqnarray*}
R(\vm_+, -2\pi) R(\vm_-, -2\pi)R(\vx, \theta) 
\end{eqnarray*}
is constructed. 
Let us examine the condition  \mbox{$-\theta (\sigma_x/2)=A_+ + A_-$}
for using the Trotter-Suzuki formula. 
By taking into account that $|\vx \rangle$ is the cyclic state for 
both $R(\vx, \theta)$ and $R(\vm_\pm, -2\pi)$, 
\begin{eqnarray*}
0&=&\langle \vx |\theta \frac{\sigma_x}{2} + A_+ + A_- |\vx \rangle
\nonumber  \\
&=& \langle \vx |\theta \frac{\sigma_x}{2}|\vx \rangle
+ \langle \vx |R(\vx, -\theta)  A_- R(\vx,\theta) |\vx \rangle
\nonumber \\
&+& \langle \vx |R(\vx, -\theta) R(\vm_-, 2\pi)  A_+ 
R(\vm_-, -2\pi)R(\vx,\theta) |\vx \rangle . 
\end{eqnarray*}
In other words, the condition  for using the Trotter-Suzuki formula 
\mbox{($-\theta (\sigma_x/2)=A_+ + A_-$)}
is equivalent that the sum of dynamic phases during the composite 
quantum gate is vanishing.

We show that all GQG's 
with Aharanov-Anandan phases are fully compensating 
composite quantum gates that are robust 
against control field strength errors.
We also show 
\mbox{(1) how} SCROFULOUS pulses and W1 sequences~\cite{cp_qip}
are geometrically constructed and 
\mbox{(2) the} condition for using the Trotter-Suzuki 
formulas~\cite{BHC2004,TS_formula}
is equivalent that the sum of dynamic phases in a 
composite quantum gate is vanishing.
Although the discussed case is not general, our observation provides 
a physical view on the Trotter-Suzuki formulas. 
In conclusion, 
{\it (no)dynamic phases are important to obtain reliable quantum gates. }

We would like to thank Mikio Nakahara, Yukihiro Ota, and Tsubasa 
Ichikawa for discussions. 
This work was supported by ``Open Research Center'' Project for
Private Universities: Matching fund subsidy from Ministry of
Education, Culture, Sports, Science and Technology.


\begin{thebibliography}{99}
\bibitem{memo}
We often take $T=1$ and then omit $T$ in Eq.~(\ref{eq:Hamiltonian}) 
for readability.
\bibitem{comment1}
The condition on the dynamic phases is 
\begin{eqnarray*}
&&2\theta_1\cos\phi_1+\pi\cos\phi_1\cos(\phi_1-\phi_2)\\
&+&\pi \cos \theta_1 \sin \phi_1 \sin (\phi_1-\phi_2) 
=0
\end{eqnarray*}
by employing Eq.~(\ref{eq:dyn_p}). From Eq.~(\ref{eq:SCR1}) 
and this condition, we get Eq.~(\ref{eq:SCR3}).
\bibitem{my_cal}
We start from 
\mbox{$\cos^2(\phi_1-\phi_2)+\sin^2(\phi_1-\phi_2)=1$} and substitute 
Eqs.~(\ref{eq:SCR1}-\ref{eq:SCR3}) into it. Then, we obtain 
\begin{eqnarray*}
\frac{1}{\theta_1}\sin\theta_1 = \frac{2}{\pi}\cos(\theta/2),
\end{eqnarray*}
and 
\begin{eqnarray*}
\phi_1&=& {\rm arccos}\left( \frac{-\pi \cos \theta_1}
{2\theta_1\sin(\theta/2)}\right),\\
\phi_2&=&\phi_1-{\rm arccos}\left(\frac{-\pi}{2\theta_1}\right).
\end{eqnarray*}
\end{thebibliography}
\end{document}